\documentclass[preprint,aps]{revtex4}
\usepackage{amssymb} \usepackage{txfonts} \usepackage{graphicx}
\usepackage{dcolumn} \usepackage{bm} \usepackage{epstopdf}

\begin{document}
\makeatletter
\makeatother

\title{Bending a periodically layered structure for transformation acoustics}
\author{Zixian Liang and Jensen Li}
\email{jensen.li@cityu.edu.hk}\affiliation{Department of Physics and Materials Science, City University of Hong Kong, Tat Chee Avenue, Kowloon, Hong Kong, China} 

\begin{abstract}
Anisotropic acoustic metamaterials have been proved very useful for their high potential in guiding and manipulating sound energy. In this paper, we further develop the idea by using periodically layered structures for transformational acoustics. Such a simple scheme periodically inserts identically bent solid plates in a background fluid. It forms a metamaterial with high refractive index normal to the curved plates and an index near to one along the plates. We show that the periodically layered structure, combined with transformation approach, can be cut into particular device shapes for acoustic cloaking and illusion.
\end{abstract}
\maketitle

Metamaterials are bringing new opportunities in manipulating electromagnetic, acoustic, and water waves, particularly through the framework of transformation optics \cite{Pendry2006, Leonhardt2006, SchurigD2006, Cummer2007njp, ChenH2007apl, GreenleafPRL2008, FarhatPRL2008, Norris2009, FarhatPRL2009, FangPRL2011}. In acoustics, we can design such metamaterials which consist of artificial periodic subwavelength structures to make their description using effective densities and moduli valid. Moreover, due to the engineering flexibility associated with these artificial materials, extreme values in their constitutive parameters can be obtained by using resonating subwavelength units \cite{liu2001, Fang2006, ShengPRL2008, KimPRL2010}. Nevertheless, there is always a tradeoff between extreme constitutive parameters near resonance and a large working frequency bandwidth with low loss. It is therefore reasonable to investigate moderate acoustic metamaterials which can work in broadband with low loss. By introducing asymmetry in the shape of unit cells or in the constructing units, we can create acoustic metamaterials with broadband anisotropy \cite{Torrent2008b, Jensen2008NJP}. In particular, a one-dimensional lattice of alternating layers of a solid and a fluid has been found to possess large anisotropy. Such a simple periodically layered structure has been used to guide sound energy along the direction of fluid perforation for different applications. Acoustic superlenses, hyperlenses, acoustic devices with extraordinary transmission and recently an acoustic cavity have been fabricated utilizing this guiding capabilty of anisotropic acoustic metamaterials \cite{Cai2007, Lu2007, ChristensenPRL2008, JensenLi2009, HJia2010,  AmirkhiziWM2010, Torrent2010, JZhu752}. 

\begin{figure}
\centering
\includegraphics[width=0.49\textwidth]{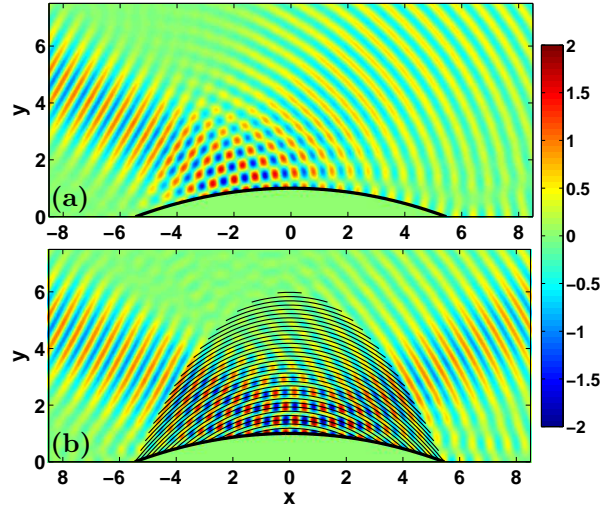}
\caption{\label{fig:carpetcloak}(a) Pressure field pattern when only the object (the bottom curved reflecting hard surface) is present without the cloak. (b) Pressure field pattern with the microstructured carpet cloak. The curved black lines represent the inserted brass plates into water. All length scales are normalized to $h_1$. The period of the curved structure is $0.15h_1$, and the filling ratio of brass is $0.074$. A Gaussian beam of width $3.5h_1$ at 60 degrees impinges on the surface from the left at a wavelength of $0.58h_1$ in background water. In the simulation, water/brass has a density of $1000 kgm^{-3}/8500 kgm^{-3}$ and longitudinal sound speed of $1500 ms^{-1}/4700 ms^{-1}$.}
\end{figure}

In this work, we further extend the usage of such a simple periodically layered structure into the domain of transformation acoustics. In particular, we bend a periodic stack of planar solid plates with a percolating fluid. The periodically layered structure can be used to guide sound waves as a transformational acoustic device through cutting a particular shape out of the periodic medium. Instead of the isotropic multilayer approach in constructing cylindrical cloaks \cite{LiuXJAPL2008} with the isotropic constitutive parameters varying in space, we consider only two types of constant materials: a solid and a fluid with constant material parameters and constant filling ratio. Let us begin with the planar solid plates, with volume filling fraction $\eta$ in a fluid, stacking along the $y$-direction with a subwavelength lattice constant. Such a metamaterial (with large density contrast between the fluid and the solid) can be represented by an effective fluid medium with 
\begin{equation}
\label{eq:effrhoalpha}
\frac{1}{\rho_\alpha}=\frac{\eta}{\rho_{1}}+\frac{1-\eta}{\rho_{0}}
\approx\frac{1-\eta}{\rho_{0}},\  
{\rho_\beta}=\eta \rho_{1}+(1-\eta)\rho_{0},
\end{equation}
where $\rho_\alpha$/$\rho_\beta$ is the effective density in a direction parallel/normal to the plates, and the effective bulk modulus $B$ governed by
\begin{equation}
\label{eq:effB}
\frac{1}{B}=\frac{\eta}{B_{1}}+\frac{1-\eta}{B_{0}}
\approx\frac{1-\eta}{B_{0}}.
\end{equation}
\begin{figure}
\centering\includegraphics[width=0.49\textwidth]{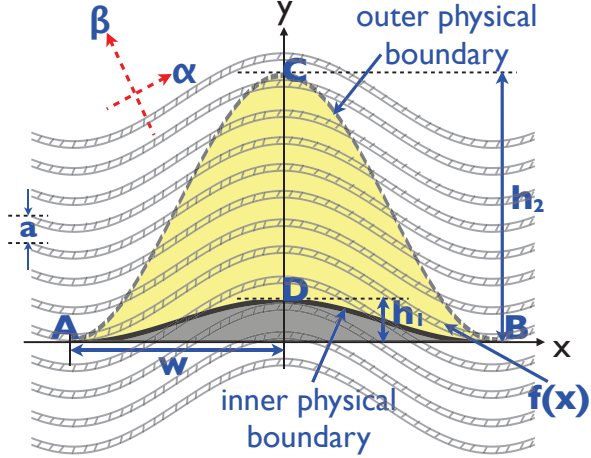}
\caption{\label{fig:schematic}Schematic of the transformation acoustical device. The device (yellow in color) can be constructed by cutting the periodically layered structure (curved solid plates with function $f(x)$ stacking periodically along the $y$-direction, shown in light grey color) into a particular shape. $\alpha$ and $\beta$ are the principal axes of the acoustic metamaterial.}
\end{figure}
Subscript $1$/$0$ indicates the densities and the bulk moduli of the solid/fluid media. The same effective medium can also be described by the refractive indices normalized to the background fluid using $n_{\alpha/\beta}=\sqrt{\rho_{\alpha/\beta}/B}/\sqrt{\rho_0/B_0}$. In the direction normal to the plates, the index $n_\beta$ is higher than one since the plates are blocking the waves in this direction, and along the plates, the index $n_\alpha$ is near to one since the fluid can percolate freely. Now, we bend the layered structure along a curve of function $f(x)$. For simplicity, we only consider wave propagation on the $x$-$y$ plane. The bent structure is our interested physical medium which is $y$-periodic with the same filling ratio $\eta$ for the solid. The effective medium is still described by Eq. (\ref{eq:effrhoalpha}) and Eq. (\ref{eq:effB}) with the principal $\alpha$ and $\beta$-axes now aligned with the tangent of $f(x)$, as shown in Fig. \ref{fig:schematic}. In fact, it can be easily proved that the bending function $f(x)$ induces a coordinate transform between the two spaces, given by
\begin{equation}
y=y'+(1-n_\alpha^2/n_\beta^2)f(x'),\ x=x',
\end{equation}
where the unprimed coordinates represent the physical medium or the bent structure while the primed coordinates represent the virtual system consisting the planar structures. It ensures the two effective media transform to each other according to the philosophy of transformation acoustics \cite{Cummer2007njp, ChenH2007apl}. The bending function $f(x)$ is assumed to be slowly varying so that the refractive indices are accurate up to the 1st order of $f'(x)$. The coordinate transform serves as a starting point to construct transformation acoustical devices. For example, if the device now transforms between a fluid of isotropic index $n_\alpha$ and the bent structure, we employ an additional stretching in the y-direction and the two media are transformed according to
\begin{equation}
\label{eq:map}
y=n_{\alpha}/n_{\beta}y'+(1-n_\alpha^2/n_\beta^2)f(x'),\ x=x'.
\end{equation}
From the transformation, the device has a physical outer boundary (which is invariant upon the transformation):
\begin{equation}
\label{eq:outbnd}
y=(1+n_\alpha/n_\beta)f(x),
\end{equation}
as shown as the curve ACB in Fig. \ref{fig:schematic}. On the other hand, we can set the virtual inner boundary to be $y'=0$ which maps to the inner physical boundary (by Eq. (\ref{eq:map})):
\begin{equation}
\label{eq:inbnd}
y=(1-n_\alpha^2/n_\beta^2)f(x),
\end{equation}
as shown as curve ADB in Fig. \ref{fig:schematic}. Such a device can act as a cloak in a background fluid of isotropic index $n_\alpha$ for an object sitting on a hard surface. It is also called an acoustic carpet cloak and several other coordinate transforms for it have been investigated \cite{JensenPRL2008, ZhaoAPL2010, StefanOE2010}. Here, we actually use the curved brass plates for the miscroscopic construction. The cloak squeezes the area between the outer boundary and the $x$-axis to the yellow region between the two physical boundaries and creates a hidden tunnel (or the object in grey color) below the inner physical boundary shown in Fig. \ref{fig:schematic}. Here, we begin from an anisotropic metamaterial with alternating planar layers of brass plates and water with a volume filling fraction of 0.074 for brass. The metamaterial gives rises to $n_\beta=1.2$ and has an advantage that $n_\alpha=0.996$ which is very near to one. Therefore, the cloak can also work fine in background water as a good approximation, a special property due to the construction scheme for our acoustic metamaterials. Next, we bend the brass plates using function $f(x)=\xi(h_1-R+\sqrt{R^2-4x^2})$, where $\xi=3.21$, $R=60h_1$. We can now cut such a medium into the right shape according to Eq. (\ref{eq:outbnd}) and Eq. (\ref{eq:inbnd}), and $h_1$ is the height of the hidden tunnel for the cloak. In essence, the microstructured cloak is constructed just by inserting identically bent brass plates periodically into the background water. It significantly simplifies the microstructured design and fabrication of an acoustic carpet cloak with a general shape for the inner boundary. To test the effectiveness of the design, we have employed full wave simulation (COMSOL Multiphysics) by launching a Gaussian beam at 60 degrees towards the cloak. The wavelength in water is set as $0.58h_1$. The total pressure field pattern with only the curved reflecting surface without the presence of the cloak is shown in Fig. \ref{fig:carpetcloak}(a). 
The curved reflecting surface scatters the incident beam into a broad and curved wavefronts. Now, the situation with the presence of the cloak (periodic curved brass plates) is also simulated. The curved brass plates are outlined as thin black lies in Fig. \ref{fig:carpetcloak}(b). The periodicity is chosen as 0.15$h_1$, a value few times smaller than the wavelength in background water. The corresponding pressure field is shown in Fig. \ref{fig:carpetcloak}(b). On the whole, the cloak cancels out the scattering of the object in a way that the cloak with the object just looks like a flat hard surface in reflecting the incident Gaussian beam also at 60 degrees without additional scattering. Moreover, the field inside the cloak shows interference pattern between the incident and the reflected beam. The interference pattern is simply squeezed upwards, comparing to the field when only a flat hard surface is present. In our construction scheme, we have actually lumped the bulk modulus into the density tensor by $\rho_\alpha \to \rho_\alpha B_{0}/B$, $\rho_\beta \to \rho_\beta B_{0}/B$ and $B \to B_{0}$ because the bulk modulus given by our acoustic metamaterial is very near to the one of background water. This is called the reduced parameter approximation. It has the advantage that we can concentrate on constructing the density tensor from acoustic metamaterials without considering the bulk modulus. Although there is such an approximation (which introduces an impedance mismatch between the background fluid and the cloak), there is only little spurious reflections from the boundary of the cloak in the simulation. 

\begin{figure}
\centering
\includegraphics[width=0.49\textwidth]{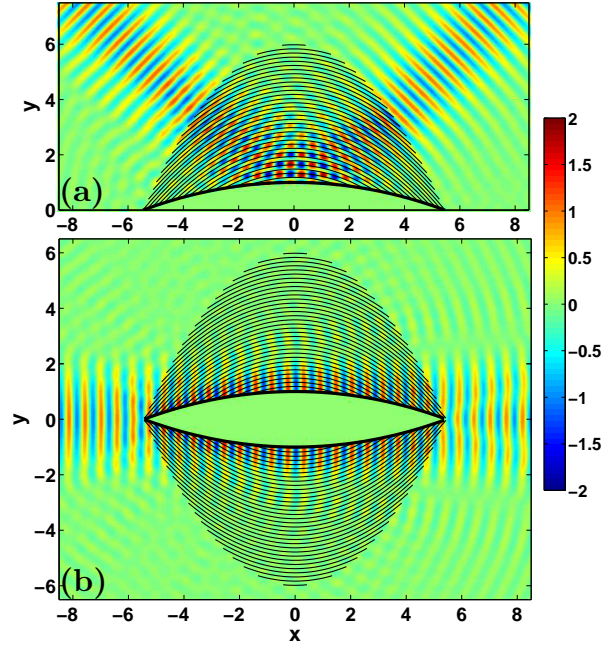}
\caption{\label{fig:otherangles}(a) Pressure field pattern for a Gaussian beam incident at 45 degrees to the same microstructured carpet cloak in Fig. \ref{fig:carpetcloak}(b). (b) Pressure field pattern of the beam launching horizontally to the mirrored carpet cloak.}
\end{figure}
In our analysis, the refractive indices induced from the transformation (Eq. (\ref{eq:map})) are regarded as constants in magnitudes (only the directions of the principal axes are changing) as an approximation in constructing the transformation devices using the periodic curved solid plates. Technically speaking, the magnitudes of the indices induced from the transformation (Eq. (\ref{eq:map})) should vary along the x-direction, related to the slope of the curved layered structure $f'(x)$. The approximation becomes less accurate for a larger slope. Fig. \ref{fig:otherangles} shows the simulations of the cloak at other two different incident angles. Fig. \ref{fig:otherangles} (a) depicts the pressure field pattern of the Gaussian beam incident at $45$ degrees to the cloak. The cloak cancels out the scattering of the object and the beam is reflected at 45 degrees as expected and the approximation is accurate. However, when the Gaussian beam is launched at 90 degrees (horizontally) to the cloak composed of two symmetric carpet cloaks, as shown in Fig \ref{fig:otherangles} (b), there is a little bit scattering at the corners of the cloak. For the ideal case, the cloak compresses the object to a very thin hard plate parallel to the beam direction and we should expect the beam pass through the cloak without any scattering. Now, because $f'(x)$ at the corners of the cloak is largest, the approximation becomes less accurate. 

\begin{figure}
\centering
\includegraphics[width=0.49\textwidth]{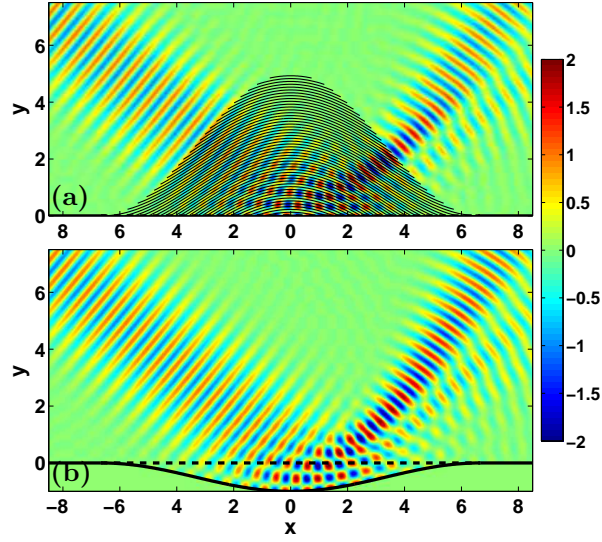}
\caption{\label{fig:holeillusion}Pressure field pattern with (a) the trough illusion device putting on a flat hard surface and (b) the equivalent water trough (a concave reflecting hard surface of height $h_1$.  All length scales are normalized to $h_1$). The period of the curved brass plates is $0.1h_1$, and the filling ratio of brass is $0.074$. A Gaussian beam of width $4h_1$ at 50 degrees impinges on the surface from the left at a wavelength of $0.58h_1$ in background water.}
\end{figure}
Apart from a cloak, we can also use the developed periodic curved structure to guide wave in order to give other kinds of acoustic illusion. For example, we can design a device sitting on a flat hard surface so that an observer gets a trough illusion that a trough of certain depth below the hard surface exists. In this example, we use the same anisotropic metamaterial with another bending function $f(x)=\xi h_1cos(\frac{\pi x}{2 w})^2$, where $\xi=2.67$ and $w=6.7h_1$. The outer boundary of the device is still governed by Eq. (\ref{eq:outbnd}) but the inner physical boundary is now given by $y=0$ which maps to the virtual inner boundary $y'=-h_1cos(\frac{\pi x'}{2 w})^2$ according to Eq. (\ref{eq:map}), where $h_1$ is the perceived depth of the trough. In this case, we have set the periodicity to be $0.1h_1$ and the wavelength in background water to be $0.58h_1$. The device transforms the incident beam as shown in Fig. \ref{fig:holeillusion} (a). The pressure field pattern outside the device is very similar to the case that a trough is dug into the flat surface, as shown in Fig. \ref{fig:holeillusion} (b). There is scattering induced by the trough and a parasitic weaker beam exists.

In conclusion, we have investigated a scheme in constructing transformation media simply by periodically inserting bent solid plates into a background fluid. It allows us to guide sound waves just by controlling the directions of the principal axes of the equivalent anisotropic metamaterial. Within the framework of transformation acoustics, we have used the scheme to achieve acoustic carpet cloaking and trough illusion by cutting the periodically layered structure into particular shapes.

Acknowledgement

JL thanks for the support from City University of Hong Kong SRG grant number 7002598.

\end{document}